# Broadband dielectric characterization systems for food materials

J. K. Hamilton[1,2,*], C. P. Gallagher[2], C. R. Lawrence[1], and J. R. Bows[3]

[1]*QinetiQ Ltd, Cody Technology Park, Ively Rd, Farnborough, GU14 0LX*
[2]*Department of Physics and Astronomy, University of Exeter, Exeter, Devon, EX4 4QL, UK*
[3]*PepsiCo, Leicester, LE4 1ET*

*[jkhamilton@QinetiQ.com](jkhamilton@QinetiQ.com)

**Abstract:** A detailed knowledge of the dielectric properties of food materials is vital to any electromagnetic-based treatments, ranging from reheating meals in a domestic microwave oven through to sterilization processes. The uniformity and rate of heating of the food is highly dependent upon the dielectric constant and dielectric loss of the food material, with both parameters being intrinsically temperature-dependent. In this work, we explore various methods for the dielectric characterisation of materials over a wide frequency range (1 MHz to 20 GHz), to cover RF and microwave frequencies, as well as understand the impact of ingredients (e.g. salt, sugar) and temperature on free and bound water resonances. Such a large frequency range gives information that is invaluable when designing future equipment that relies on multi-frequency/broadband microwave heating techniques. The characterization methods of interest are the open-ended coaxial dielectric probe, the broadband dielectric broadband spectrometer, and a custom stripline resonator. For ease of comparison, a food standard is used – instant mashed potato – which allows for a large set of samples to be measured – with various moisture contents. When directly compared at 1 GHz, all three methods produced comparable dielectric properties for 20% and 30% potato mixtures.

## Introduction

Electromagnetic heating is a commonly used tool in the food industry (e.g. bacon cooking, tempering, pasta drying, pasteurization). As is the case for any other dielectric, the heating of foodstuffs via electromagnetic waves is primarily determined by the dielectric properties of the irradiated material. The dielectric constant ($\varepsilon'$) reflects the ability of a material to store electromagnetic energy, whereas the dielectric loss ($\varepsilon''$) measures the ability of a material to dissipate electromagnetic energy as heat. The electromagnetic properties of a material affect the heating uniformity, rate of heating and penetration depth of energy into the food. As a result, the effectiveness of heating when using electromagnetic methods depends strongly on a clear understanding of the electromagnetic properties of various food materials.

The majority of food materials contain a certain percentage of water – which can be described as the moisture content. Due to its presence, it is important to understand the dielectric properties of water and its effects on food materials. Deionised water has a relatively high dielectric constant of 78.4 (at frequencies < 1 GHz) and will decrease as the frequency is increased. This frequency-dependent response is due to dielectric polarisation under the influence of an electric field and the polar nature of the water molecules. The dielectric loss of water can range between 5 and 40 between 1 GHz and 20 GHz. This change in dielectric loss – as a function of frequency – can have a drastic effect on the heating rate of water, and hence a food material. The use of accurate dielectric characterisation data has been shown to be incredibly important for moisture monitoring and crop drying [1-2]. It should be stressed that the materials of interest for the food industry and agricultural sector are not solely affected by the moisture content, as they are typically complex mixtures of various ionic solutions, including those based upon salts, starches and sugars. The typically used frequencies within the food industry are 13.56 MHz, 27.12 MHz, 915 MHz, and 2450 MHz [3].

Due to the importance of understanding these properties for electromagnetic-based food preparation and sterilization processes, there has been a vast amount of published research concerning measurement of the dielectric properties of various food materials [4-9]. A food standard (or model food) is desirable, enabling relatively flexible control over the moisture, salt and sugar content. Mashed potato is commonly employed as a food standard due to its homogeneity and the ability to easily vary its moisture and salt content, allowing the dielectric properties of many food materials to be replicated [10-14]. By using instant potato powder, a parametric sweep for the moisture of the mixture can be performed. It is worth noting that Smash® has a base salt content of 0.17% by weight (Smash®, Premier Foods).

The aim of this work is to explore the use of three different systems that are typically used for dielectric material characterization, combining them in order to obtain data across the widest possible frequency bandwidth. The first is the open-ended coaxial dielectric probe, which is the conventional method for measuring liquid samples between 100 MHz and 20 GHz [15, 16]. The second is the dielectric broadband spectrometer, which is a capacitance measurement typically used for characterising solid dielectric materials between 1 MHz and 1 GHz. Finally, the third is a stripline probing method that was developed in-house, with an operating range of 100 MHz to 20 GHz. The motivation behind by characterizing food materials over such a broad frequency and moisture range, is that additional information could be provided compared to only investigating the conventional frequencies used in the food industry. This could give motivation for future equipment to utilizing additional frequencies dependent on the moisture content and temperature of the food material.

**Explored methods and results**

In this work, various mixtures were investigated to enable an exploration of the effects of moisture content. The samples were produced by mixing different proportions of the components (by weight). The instant mashed potato powder was added to deionised water and mixed for a minimum of 30 seconds or until homogeneous. In total, 8 potato mixtures were created to explore a full range of moisture contents (20% powder by weight (wt), 30% wt, 40% wt, 50% wt, 60% wt, 65% wt, 70% wt and 75% wt). Each mixture was made prior to a measurement; this is to reduce variations in moisture content over time. Depending on the method of dielectric characterisation employed, repeat measurements were conducted by using small volumes of the mixture from each batch of the desired moisture content. The limiting factors for the sample volumes depended on the measurement technique, for example, the stripline and Novocontrol systems have a finite sample cell size – discussed later. In comparison, the dielectric probe does not limit the sample volume, however, this method only probes the material a few millimeters into the surface.

**The open-ended coaxial dielectric probe**

The conventional method used for determining the dielectric properties of a foodstuff is typically the open-ended coaxial dielectric probe [16, 17]. This system (Figure 1) consists of a truncated section of a transmission line. The electromagnetic field propagates along the coaxial line and reflection occurs when the electromagnetic field encounters an impedance mismatch between the probe and the sample's surface. This method is commonly used for the characterisation of liquid and liquid-like materials, with the probe completely inserted into the material and the surface dielectric response measured over a typical range of 0.2 GHz to 20 GHz at a high precision. The open-ended coaxial method assumes a homogeneous sample that is in good contact with the probe. Therefore, limitations of this method include possible variations in measurements due to pressure expressing water from the sample, as well as the formation of air bubbles on the end of the probe. Another possible limitation is that this method only measures the dielectric properties of the surface of the material ($\sim$ 2 mm into the material).

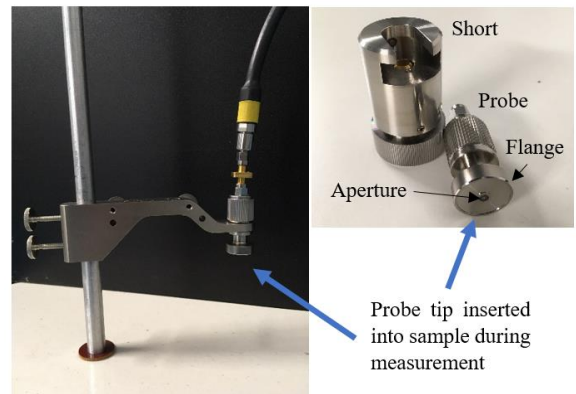

**Figure 1.** Photograph of the open-ended coaxial dielectric probe connected to a Vector Network Analyser (VNA).

For this method, the coaxial probe is calibrated to a short, deionised water and free-space laboratory conditions (temperature 21°C). In an attempt to eliminate variation across the surface of the material, measurements were conducted at various locations on the surface of the sample and averaged. Figure 2 shows the real and imaginary components of the permittivity for all mixtures with 0.2 GHz to 20 GHz.

When the moisture content is the lowest (75% wt powder), the dielectric properties are comparable to those of dried instant mash potato (without any rehydration). As the moisture content is increased, the dielectric constant increases over the investigated frequency range until 20% wt. For this case, at the lowest frequencies < 0.4 GHz the increase has reduced – falling between the response from 40% wt and 30% wt. The trend of dielectric constant of the 20% wt mixture shows a trend similar to that of pure water – tending towards a plateau at the low frequency end, as shown in Figure 2c. This is due to the high moisture content resulting in the mixture being similar to the dielectric decrement effect seen for saline solutions [17-18]. This effect is due to an effective shielding due to the ions present within the solution. An example of this is when the sodium and chloride ions – within a saline solution – dissociate in the solution, resulting in an electric field between them which the polar water molecules tend to align with. This creates a region that effectively shields the core of the sample and lowers the water molecules' response to the external field. Similar to the dielectric constant, the dielectric loss has a typical trend that increases with increasing moisture content across the investigated frequency range. However, once again there is a decrease once the moisture content is at its highest value (20% wt). For this mixture, the dielectric loss is similar to that seen for the 50% wt mixture between 0.1 GHz to 2 GHz then increases at the frequency increases.

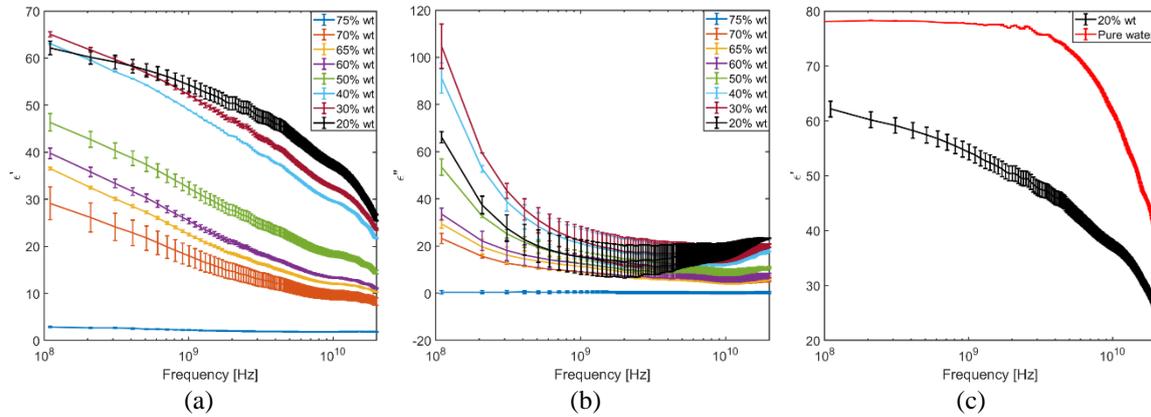

**Figure 2.** Dielectric properties for a range of mashed potato mixtures, using the open-ended dielectric probe method. (a) The dielectric constant and (b) the dielectric loss. The associated errors arise from repeat measurements across the surface of the mixture. (c) Comparison between the dielectric constant of pure water and the 20% wt mixture. Measured at laboratory conditions (temperature 21°C).

**Novocontrol dielectric broadband spectrometer**
The Novocontrol dielectric broadband spectrometer (Figure 3) measures the impedance spectrum $Z^*(f)$ of a sample material placed between two electrodes. The bulk intrinsic electric material's properties such as the complex permittivity, stated as

$$\varepsilon^*(f) = \frac{C^*(f)}{C_0}, \qquad (1)$$

are evaluated from the complex sample capacity,

$$C^*(f) = \frac{1}{i2\pi f Z^*(f)}, \qquad (2)$$

$$C_0 = \frac{\varepsilon_0 A}{d}, \qquad (3)$$

where $A$ is the electrode area and $d$ is the sample thickness. These measurements can be conducted automatically over an ultra-broadband frequency range: 1 Hz to 1 GHz (9 decades) [18]. For ease of comparison, this work will focus on the 1 MHz to 1 GHz range.

To accommodate the liquid-like structure of the instant mashed potato, a sample holder was designed and fabricated from PTFE. The sample holder was built to allow for a sample volume of approximately 150 mm$^3$. The holder's presence was compensated for by measuring the empty pod and subtracting the capacity of air (0.146 pF). This calibration allows the food material's response to be inferred, rather than the combined response of food material and PTFE holder. Once filled with the instant mashed potato, the food material forms a tight seal to the electrodes, minimizing air bubbles at the surfaces. The measurement set up is calibrated to eliminate errors – such as stray capacitance. The measurements are then verified with measurements standards of known permittivity for example: water, PTFE, and air. The electrodes have a possibility of creating artefacts at long timescale i.e. much lower frequencies than that are presented in this work. This can cause issues when the materials are highly conductive or polarized, water is an example of such a material. However, such artefacts can be calibrated and removed from the material data [19], this is something that needs to be consider when using this method to characterize the dielectric properties for frequencies < 1 KHz.

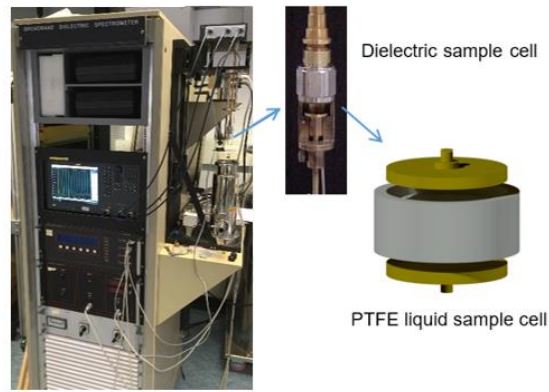

**Figure 3.** Photograph of Novocontrol broadband spectrometer and dielectric sample cell. The liquid sample pod (shown as a schematic) is placed inside the dielectric sample cell during measurements.

Using the broadband spectrometer, the dielectric properties were swept from 1 MHz to 1 GHz at room temperature, with each data point being averaged over 20 measurements in a single automated sweep. This investigation will focus on the higher moisture content samples: 50% wt powder to 20% wt powder. Figure 4 shows the measured dielectric constant and dielectric loss for mixtures of 50% wt to 20% wt. For each mixture, the dielectric constant slightly increases for increasing moisture content, until the 20% wt mixture where – similarly to the open-ended dielectric probe – there is a plateau between 10 MHz to 1 GHz and the values drop below the lower moisture content mixtures (namely 30% wt and 40% wt). The dielectric loss increases approximately linearly with increasing moisture content until the limit of the 20% wt mixture. The values of both the dielectric constant and the dielectric loss have a drastic dependence on frequency.

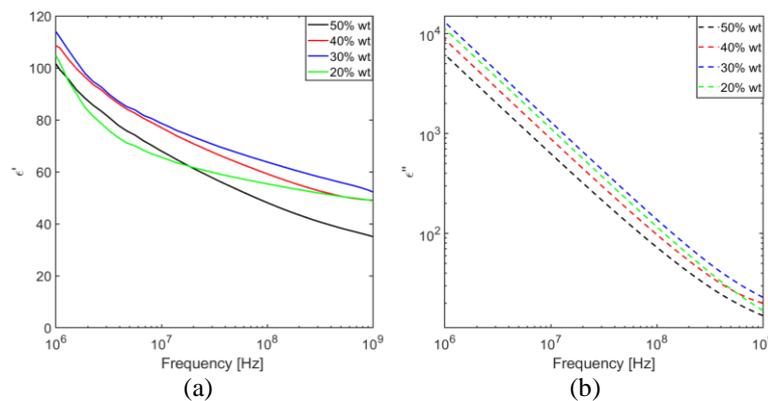

**Figure 4.** Dielectric properties for a range of mashed potato mixtures, using the broadband dielectric spectrometer. (a) The dielectric constant and (b) the dielectric loss. Measured at laboratory conditions (temperature 21°C).

One of the key advantage of the broadband dielectric spectrometer is the capacity to run automated temperature sweeps. The temperature dependence of dielectric properties for food materials subject to microwave heating processes is an essential understanding required for optimization of microwave devices. An understanding of the effect of temperature on the dielectric constant and loss values is not only important for dielectric heating applications, but also food sterilization (130°C) and frozen food processes (temperatures down to -40°C). To regulate the temperature during the sweep, a temperature probe is inserted into the sample cell. The whole RF extension line is inserted into the cryostat where the temperature is externally controlled by software. The sample is automatically brought to the desired temperature before the frequency sweep is preformed – with the sample held at that temperature. The temperature of the system is monitored at three key locations: the cryostat, the sample cell within 5 mm of the sample), the heating element. Once the sample cell temperature has stabilized at the desired temperature the system automatically measures the dielectric properties.

Figure 5 shows an example of the dielectric constant and dielectric loss for a 30% wt mixture over a temperature range from 0°C to 80°C. The dielectric constant is relatively constant with temperature at microwave frequencies; however, there is a sharp increase at radio frequencies. The dielectric loss typically increases with increasing temperature for all frequencies lower than 1 GHz. Figure 5c visualizes the temperature dependence for the dielectric constant and dielectric loss by taking the dielectric properties at 100 MHz, 450 MHz, and 1 GHz for the full investigated temperature range.

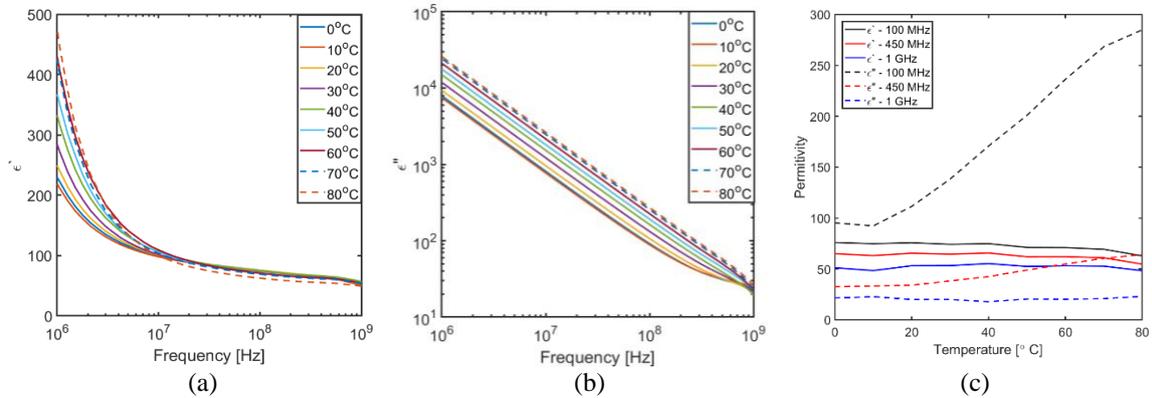

**Figure 5.** Dielectric properties for the 30% wt potato mixture, as a function of temperature. This data was measured using the broadband dielectric spectrometer to determine (a) the dielectric constant and (b) the dielectric loss. (c) Dielectric constant and dielectric loss as a function of temperature for 100 MHz, 450 MHz, and 1 GHz.

**Stripline**
Stripline measurements were taken using a brass stripline fixture with inner cavity height, $h$ = 3.00 mm, copper central conductor with thickness, $t$ = 0.15 mm, and width, $w$ = 3.83 mm. The de-embedding of S-parameter data recorded with the stripline was achieved by using a secondary set of calibration terms determined from multiple 'short' readings of the empty stripline fixture. Further details for the stripline geometry and de-embedding of S-parameter data can be found in [20]. Figure 6 shows a photograph – and schematic – of the in-house stripline system: the dielectric sample is placed in the center of the cavity and S-parameters are recorded over the desired frequency range. Samples are placed above and below the central conductor to full the cavity, it is important, that samples fit snugly into the stripline, as air gaps will cause erroneous data extraction.

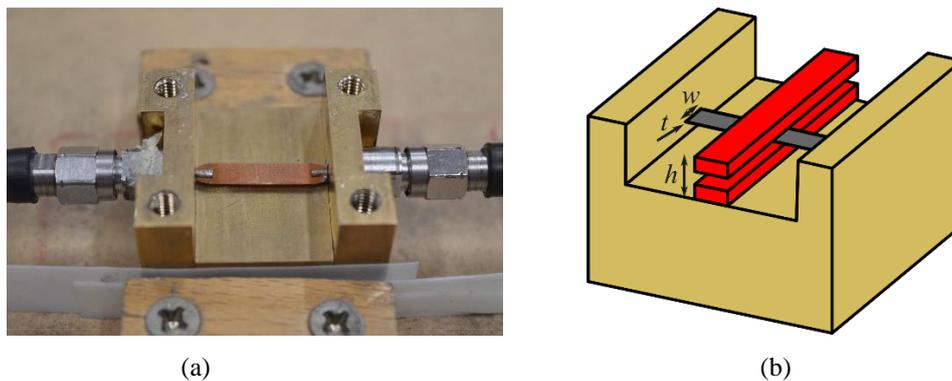

Figure 6. (a) Photograph of stripline setup. (b) Schematic of the stripline setup. The red region depicts the sample space.

The stripline method is used to characterise dielectric materials at frequencies between 0.1 GHz and 20 GHz – a similar range to that of the open-ended probe method. Figure 7 shows the dielectric response for the 20% wt (blue), 25% wt (red) and 30% wt (black) mixtures. The associated errors arise from repeat measurement of the mixtures. Figure 7a shows a sharp increase in dielectric constant between 10 GHz and 20 GHz, then tending towards a plateau as the frequency is decreased. Figure 7b shows a drastic increase in the dielectric loss at low frequency (< 1 GHz). The trends in the frequency dependence shown in this data matches those previously shown for the open-end dielectric probe and broadband spectrometer methods. Interestingly, these results from the stripline method suggest that the variation in dielectric response between 20% wt and 30% wt is within the uncertainty of the measurement. This implies that the moisture content – of this range – has little effect on the dielectric response, a result that matches what has previously been shown by Guan *et al* [15]. However, the other methods presented in this work would suggest that the moisture content offers significant contributions to the dielectric constant and dielectric loss.

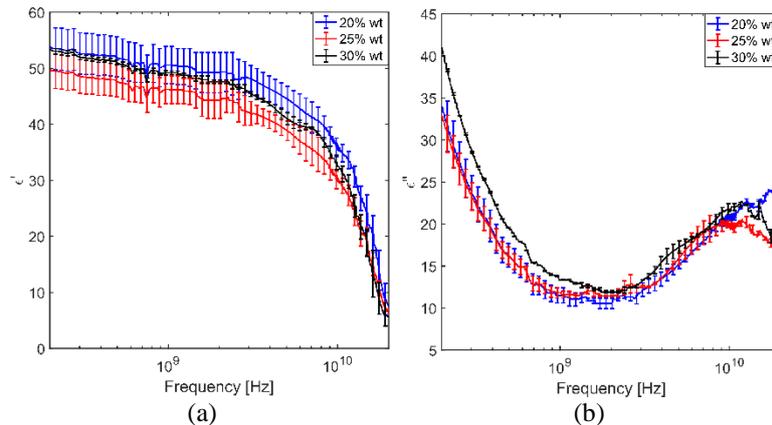

**Figure 7.** Dielectric properties for a range of mashed potato mixtures, using the stripline method to measure (a) the dielectric constant and (b) the dielectric loss. The associated errors arise from repeat measurement. Measured at laboratory conditions (temperature 21°C).

**Comparison**
The three methods have an overlapping range between 0.1 GHz and 1 GHz. A direct comparison between the three methods has been conducted, with the results shown in Figure 8. Figure 8a and Figure 8b show the dielectric constant and dielectric loss for the overlapping range between 0.1 GHz and 1 GHz for the 20% wt and 30% wt mixtures. The error bars overlap at 1 GHz but the values start to diverge at lower frequencies (<0.5 GHz): this is likely due to the open-ended dielectric probe and stripline methods nearing their respective frequency limits of 0.2 GHz and 0.1 GHz. The limit is present due to the frequency limitations of the Vector Network Analyser (VNA) rather than the measurement techniques themselves. The Novocontrol and the dielectric probe results show similar values across the investigated band for both the dielectric constant and the dielectric loss, whereas the stripline data is significantly lower. The slight differences may also be due to mixture techniques and laboratory conditions. To help further compare the three processes, a direct comparison between the dielectric properties of various mixtures at a single frequency can be conducted – and is shown in Figure 9.

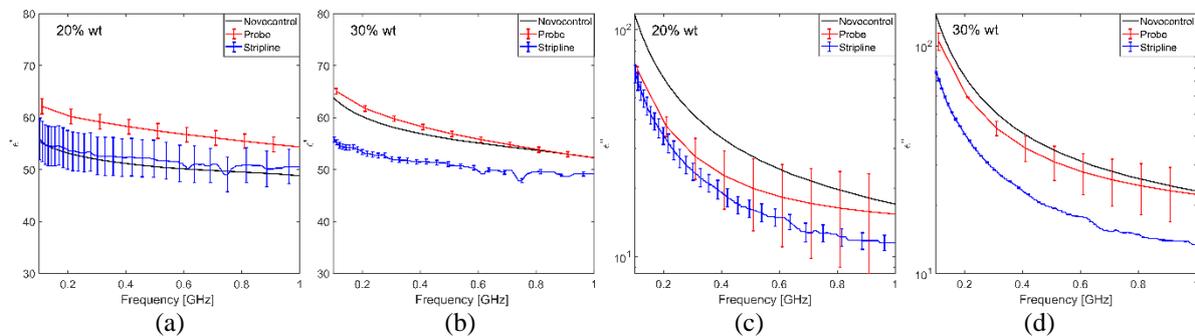

**Figure 8.** Frequency dependence comparison of the three measurement techniques: Novocontrol (black lines), open-ended dielectric probe (red lines), and stripline (blue lines). The overlapping investigated range shown is 0.1 GHz to 1 GHz. The dielectric constant is shown for (a) 20% wt and (b) 30% wt. The dielectric loss is shown for (c) 20% wt and (d) 30% wt. Measured at laboratory conditions (temperature 21°C).

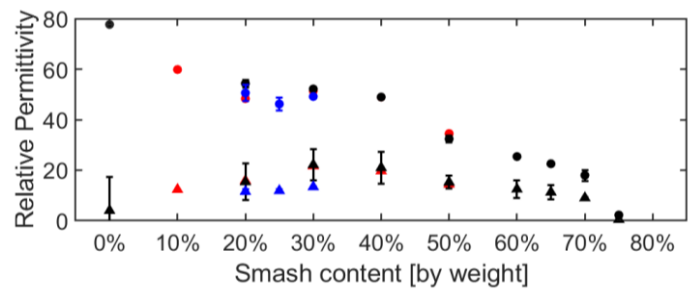

**Figure 9.** Moisture content dependence comparison for all three measurement techniques. The dielectric constant at 1 GHz for each mixture is shown as circles, and the dielectric loss at 1 GHz is shown as triangles. The colors depict the broadband dielectric spectrometer (red), open-ended dielectric probe (black) and stripline (blue) datasets. Pure water data is also shown (0% smash) – measured using the open-ended dielectric probe.

Figure 9 clearly shows that – at 1 GHz, well within the frequency range of operation for all three methods – the techniques produce similar values of both the dielectric constant and dielectric loss, for 20% wt and 30% wt. Additional information regarding the effects of moisture content are shown in this figure. The dielectric constant reduces approximately linearly with increasing potato content (and hence reduced moisture content): for example, $\epsilon = 60$ at 10% wt and $\epsilon = 20$ at 70% wt, whereas the dielectric loss is relatively constant with increasing potato content. This information would be useful for microwave processing of various food materials that have differing moisture contents. Such information is also highly relevant to microwave heating processes where the moisture content of food materials will change during the process, affecting the efficiency of the heating.

**Conclusion**

In this work, three different systems that are typically used for the RF characterization of various dielectric materials (e.g. ceramics, polymers, foams) were specifically explored for measuring liquid-like materials, to explore their applicability to the food production industry. The systems – an open-ended coaxial dielectric probe, a dielectric broadband spectrometer, and a stripline probing method that was developed in-house – were used to analyze mixtures of instant mashed potato and deionized water solutions, of powder content ranging between 20% wt to 75% wt, to explore the effect of moisture content on permittivity values (and hence heating). The full frequency range data for each system was explored and the three methods were directly compared between the overlapping frequency range of 0.1 GHz to 1 GHz. It was shown that all methods gave comparable values for the investigated materials and that some methods could be considered more convenient, depending on the dielectric characterisation required.

The open-ended dielectric probe appears to be a good method for >1 GHz measurements, with the stripline being a useful tool to verify the results. The good agreement between the broadband spectrometer and the open-ended dielectric probe is of interest due to that fact that one method is a surface measurement (the dielectric probe) whilst the other is a bulk measurement (broadband spectrometer). This would suggest that – whilst the simpler probe measurements are perfectly viable - the preferred method for detailed measurements of liquid-like materials would be the broadband spectrometer: this system has the capability to characterise a material over a far wider bandwidth (1 Hz to 1 GHz) whilst simultaneously offering temperature-based measurements. However, by combining all three measurements across their full frequency ranges, it would be possible to characterise the dielectric properties of liquid-like and food materials between 1 Hz and 20 GHz.

The values of both the dielectric constant and the dielectric loss have a drastic dependence on frequency. An example is the 50% wt mixture varying from $\varepsilon' = 15$ to $100$ and $\varepsilon'' = 10$ to $10^4$ between 0.1 GHz to 20 GHz. This information is invaluable when designing novel equipment that relies on multi-frequency microwave heating techniques. Such data is an incredibly useful tool for the next generation equipment for the food industry (utilizing multi-frequency sources for optimal microwave heating or sterilization), the agricultural sector, and other industries that require an in-depth understanding of the dielectric properties of materials containing moisture. The future equipment in these sectors/industries may utilize the broadband dielectric response of food materials and adaptively select frequencies dependent on food material properties, for example moisture constant, sugar content, and salt content.

**Acknowledgements**

The authors would like to acknowledge the funding made available from PepsiCo, which enabled this work. The views expressed in this report are those of the authors and do not necessarily represent the position of policy of PepsiCo Inc. This work was also supported by the UK Engineering and Physical Sciences Research Council Prosperity Partnership, TEAM-A (EP/R004781/1).

**Author contributions**

J. K. Hamilton: Methodology, Investigation, Writing - Original Draft, Visualization. C. P. Gallagher: Methodology, Investigation, Writing - Review & Editing. C. R. Lawrence: Writing - Review & Editing, Supervision. J. R. Bows: Conceptualization, Investigation, Writing - Review & Editing, Supervision

**Declaration of competing interest**

The authors declare that they have no conflicts of interest.